\documentclass[manuscript,nonacm]{acmart}

\renewcommand\footnotetextcopyrightpermission[1]{} % removes footnote with conference information in first column
\AtBeginDocument{%
  \providecommand\BibTeX{{%
    \normalfont B\kern-0.5em{\scshape i\kern-0.25em b}\kern-0.8em\TeX}}}

\begin{document}

\title[Towards Quantifying Requirements Technical Debt for software Requirements concerning Veracity]{Towards Quantifying Requirements Technical Debt for Software Requirements concerning Veracity: A Perspective and Research Roadmap}

\author{Judith Perera}
\email{jper@aucklanduni.ac.nz}
\authornotemark[1]
\affiliation{
  \institution{School of Computer Science, The University of Auckland}
  \city{Auckland}
  \country{New Zealand}
  \postcode{1010}
}

\author{Ewan Tempero}
\email{e.tempero@auckland.ac.nz}
\affiliation{
  \institution{School of Computer Science, The University of Auckland}
  \city{Auckland}
  \country{New Zealand}
    \postcode{1010}
}

\author{Yu-Cheng Tu}
\email{yu-cheng.tu@auckland.ac.nz}
\affiliation{
  \institution{School of Computer Science, The University of Auckland}
  \city{Auckland}
  \country{New Zealand}
    \postcode{1010}
}

\author{Kelly Blincoe}
\email{k.blincoe@auckland.ac.nz}
\affiliation{
  \institution{Department of Electrical, Computer and Software Engineering, The University of Auckland}
  \city{Auckland}
  \country{New Zealand}
    \postcode{1010}
    }

\author{Matthias Galster}
\email{matthias.galster@canterbury.ac.nz}
\affiliation{
  \institution{Department of Computer Science and Software Engineering, College of Engineering, University of Canterbury}
  \city{Christchurch}
  \country{New Zealand}
    \postcode{1010}
    }

%%concise list of authors' names 
\renewcommand{\shortauthors}{Perera, et al.}
\renewcommand{\shorttitle}{Towards Quantifying Requirements TD for Software Requirements concerning Veracity}

\begin{abstract}
Software practitioners can make sub-optimal decisions concerning requirements during gathering, documenting, prioritizing, and implementing requirements as software features or architectural design decisions --- this is captured by the metaphor `Requirements Technical Debt (RTD).' In our prior work, we developed a conceptual model to understand the quantification of RTD and support its management. In this paper, we present our perspective and the vision to apply the lens of RTD to software requirements concerning veracity, i.e., requirements related to truth, trust, authenticity, and demonstrability in software-intensive systems. Our goal is to cultivate awareness of veracity as an important concern and eventually support the management of RTD for software requirements concerning veracity, what we term as `Veracity Debt,' through its quantification.
\end{abstract}

\maketitle
\thispagestyle{empty}

\section{Introduction}

Software practitioners can make sub-optimal decisions concerning requirements either deliberately or inadvertently when gathering, documenting, prioritizing, and implementing requirements as software features or architectural design decisions. Some examples are failing to capture important user needs, introducing ambiguities when documenting the requirements, inadequate documentation of requirements, sub-optimal prioritization of requirements, inadequate (partial or incorrect) implementation of requirements as features, and failing to satisfy requirements through architectural design decisions \cite{perera2023quantifying}. The consequences of such sub-optimal decisions are captured by the metaphor `Requirements Technical Debt (RTD)' \cite{perera2023quantifying}.

In our previous work, we developed a conceptual model that captures the concepts and relationships between concepts to better understand and support RTD quantification so that management decisions regarding RTD management could be made in an informed manner. In this paper, we coin the term `Veracity RTD' (in short, `Veracity Debt') to capture the consequences of sub-optimal decisions made with respect to software requirements related to veracity. We then present our perspective and the vision to apply the lens of RTD quantification to Veracity requirements to quantify Veracity Debt and thereby support informed decision-making for its management. 

`Veracity requirements,' i.e., software requirements concerning veracity, refers to a specific category of software requirements related to truth, trust, authenticity, and demonstrability in software-intensive systems \cite{10.1145/3589154}. Veracity requirements can be functional, non-functional, or both, depending on the software system under development (and the context, e.g., domain constraints, system environment, and business goals). Veracity requirements are becoming an essential type of requirements to be implemented, for example, in many software-intensive systems and data infrastructures where attributes such as accuracy, precision, trust, and truthfulness are considered necessary qualities to have \cite{castka2020technology, blincoe2023human, galster2023understanding}. 

Social and Environmental Audits (SEA) is an applicable example domain for veracity requirements. SEA helps reduce social and environmental non-compliance, enhance transparency, and improve the legitimacy of firms and their supply chains \cite{castka2020technology}. Another applicable field is AI, where the trust in decisions made by AI systems is becoming increasingly important \cite{blincoe2023human}. An everyday example is the veracity of online data, e.g., fake news, malicious rumors, fabricated reviews, fake images, and videos \cite{lozano2020veracity}. The provenance of software code is an example where veracity is important when one may want to know where a piece of code originates from, who developed it, and why \cite{galster2023understanding}. 

Sub-optimal decisions made concerning veracity requirements could lead to adverse consequences. For example, if regulatory or compliance requirements are unmet (these can be related to data such as the General Data Protection Regulation (GDPR) or domain-specific such as building regulations or organic product certification), it can lead to legal concerns and eventually lead to large cost overruns, a bad reputation, and the loss of customers. Another example is that if requirements related to the veracity of data captured in software-intensive systems are not adequately implemented, this may cause security and privacy concerns and introduce biases in AI systems. Furthermore, inadequately capturing or implementing cultural data sovereignty requirements can lead to socio-cultural concerns \cite{10.1145/3589154}. Therefore, software companies and their stakeholders can benefit from the informed management of RTD, especially regarding software requirements concerning veracity, and thereby reduce and avoid adverse consequences.

%figure decision-making

The paper is structured as follows. Section \ref{sec:background} discusses the background and related work, and Section \ref{perspective} discusses our perspective on quantifying RTD for software requirements concerning veracity. Section \ref{roadmap} presents the proposed research roadmap for our future work, and the paper is concluded in Section \ref{conclusion}.

\section{Background and Related Work}
\label{sec:background}

\subsection{Veracity}

Blincoe et al. \cite{blincoe2023human} discusses `trust' as an aspect of veracity. In their report, they summarize the literature on the trust of digital technologies from a human-centric perspective. The authors identified three main behaviors of trust: competence, benevolence, and integrity for trust in general, and that a person's own disposition to trust, as well as institutional trust, are important for trust in general, according to the literature surveyed in their report. For technology-mediated trust, they identified that contextual and intrinsic properties of trust apply and that elements of trust still focus on people and organizations involved in the trust relationship. According to the authors, for software products, trust is based on both the trust of the creators and the trust of the software product itself. 

The authors suggest that transparency, explainability, trialability, interoperability, and accuracy are important for trust formation in AI systems. They also derived ethical concerns and fairness as important for trust in AI systems. The report summarizes factors (i.e., intrinsic and extrinsic properties) influencing trust in digital technologies, including software products and AI. The intrinsic properties include, for example, accuracy and correctness, security and privacy, reliability and dependability, transparency, explainability, interpretability, trialability, and fairness. Extrinsic properties include trust of the tech creators, review of tech products, users' perceptions of how decisions are made, and accountability. 

Galster et al. \cite{galster2023understanding} discuss retrofitting `provenance' (i.e., metadata about the origin, context, or history of data) to existing software systems as a means to answer the questions associated with how data is used in software systems, how data is being collected and how the data is used to make decisions. The authors work towards understanding the industry's needs regarding provenance in software. Based on their interview results, they developed an `influence map' for provenance that describes three dimensions of provenance --- When, What, and How. The influence map also includes stakeholders: regulators, third parties, customers, and developers.

Lozano et al. \cite{GARCIALOZANO2020113132} examine the concept of 'veracity' in the context of big data and state that the term `veracity' has no agreed-upon definition in the literature. They used a list of terms to define veracity related to big data, including truth, trust, uncertainty, credibility, reliability, noise, anomalous, imprecise, and quality. Their work aimed to find approaches, methods, algorithms, and tools proposed for the automatic veracity assessment of open-source data.  

Rubin et al. \cite{rubin2013veracity} developed `the big data veracity index' in their work. According to the authors, their veracity index provides a useful way of assessing systematic variations in big data quality across datasets where there is textual information. 

Luczak-Roesch et al. \cite{10.1145/3589154} provide their perspective on veracity while reporting on the Veracity Project initiated by the New Zealand government in 2021. The author's perspective is that the concept of veracity does not capture its scope multi-directionally by framing it as data quality. Instead, the authors suggest that veracity must capture multi-directional considerations such as the sovereignty of artifacts and requirements around data sovereignty and that digital artifacts cannot be disentangled from their context from the real world or the real-world artifact which a digital artifact (e.g., a digital twin) has been created for. Therefore, the authors define veracity to be a border definition: ``truth, trust, authenticity, and demonstrability" \cite{10.1145/3589154}.

\subsection{RTD and RTD Quantification}

Ernst et al. \cite{ernst2012role} define RTD as the distance between the optimal requirements specification and the actual system implementation under domain assumptions and constraints. According to the author, RTD describes trade-offs on what requirements the software development team should prioritize. Product value is measured in satisfied requirements, and RTD is incurred when requirements are prioritized in a way they do not deliver the most value to the customer \cite{ernst2012role}. They developed a tool called RE-KOMBINE to compare the current implementation to the new proposed implementation and measure the distance between the proposed solutions and the current implementation as RTD. 

Lenarduzzi and Fucci \cite{lenarduzzi2019towards} describe three types of RTD based on Incomplete Users' needs, Requirement Smells, and Mismatch Implementation in their paper. Their definition extends the definition initially proposed by Ernst et al. \cite{ernst2012role} to include upstream activities such as eliciting requirements and specifying requirements. 

According to Perera et al., \cite{perera2023quantifying}, Requirements Technical Debt (RTD) captures the consequence of sub-optimal decisions made concerning requirements during their gathering, formalization, and implementation \cite{perera2023quantifying}. The authors developed a conceptual model to capture the concepts and relationships relevant to the quantification of RTD based on their Systematic Mapping Study (SMS) and prior work about quantifying software code-related TD (e.g., Code TD, Design TD, and Architectural TD) \cite{perera2023_quantifyingTD}. Their conceptual model helps researchers and practitioners understand the quantification of RTD, envisioning that the quantification of costs and benefits associated with RTD will lead to better management of RTD. 

Their model showed that RTD has its own components, although RTD is similar in most aspects to software code-related TD. Examples of aspects unique to RTD were, Requirements Engineering (RE) costs associated with RTD and Requirements artifacts involving the user in a feedback loop different from code-related TD. Furthermore, they discussed that RTD can occur regardless of the presence of code-related TD and that rectifying or remediating RTD can also accrue benefits similar to refactoring the software code. 

%'RTD Interest' is an important concept captured in the model that is equivalent to 'Interest' paid on Financial Debt in real life. Interest can be accumulated in the form of rework or new work in the Requirements Engineering, System Design and Implementation phases in a software project. For example, as a consequence of incurring Veracity Debt, some parts of the system may require a redesign or a re- implementation leading to the consumption of more resources and delayed development.

Ojameruaye et al. \cite{ojameruaye2016sustainability} propose an economics-driven approach to evaluate sustainability requirements. The authors draw a link between architectural design decisions (i.e., strategies) and sustainability requirements, considering that software architecture is a set of design decisions that help meet functional requirements while optimizing desired quality attributes. They consider that RTD is incurred when requirements that do not deliver the most value to the customer are prioritized, aligning with the initial definition of RTD by Ernst et al. \cite{ernst2012role}. The interest on the debt is the rate of increase in the distance. Their method attempts to quantify the gap between the `level of sustainability' that will be achieved with a specific architecture and an ideal environment where sustainability requirements are entirely fulfilled (i.e., the desired level of sustainability). Sustainability requirements are considered Architecturally Significant Requirements (ASRs) in their paper. The authors extend the architecture evaluation method CBAM \cite{919103}, a scenario-based method for analyzing costs, benefits, and schedule implications of architectural decisions which is based on the Architecture Trade-off Analysis Method (ATAM) \cite{kazman1998architecture} and integrates principles of Modern Portfolio Theory (MPT) \cite{mangram2013simplified} to select the right design decision or strategies based on their impacts on different sustainability goals. The method allows developers and analysts to estimate the value of prioritized ASRs or architectures systematically. An optimal portfolio is a set of architectural design decisions that minimize costs, reduce risks, and yield the maximum benefit regarding the sustainability dimensions: technical, environmental, social, economic, and human. 

\subsection{An Illustrative Example of Veracity Requirements from the New Zealand Organic Products Supply Chain}

%\subsubsection{The New Zealand Organic Products Supply Chain Use Case}

The organics domain in New Zealand encompasses a wide variety of products, including orchards, livestock, aquaculture, honey and bee products, wild and natural products, viticulture, and winemaking \cite{NZ_organic}. The Organic Supply Chain comprises eight stages, which include inputs for organics, farming and growing, transport, processing, packaging, warehousing, shipping, and retail. Multiple stakeholders are involved in these stages, such as primary producers (i.e., farmers or growers), non-primary producers (i.e., manufacturers, distributors, exporters), retailers, and consumers. Primary and non-primary producers have to comply with regulations and standards and adhere to certain processes to obtain and maintain the organic certification each year. The supply chain operations and the organic produce are certified by Certification Agencies. Audits are carried out by the certification agencies (either informed or uninformed) to ensure that the veracity and integrity of the products and the supply chain process are maintained throughout.

Our engagements with stakeholders in the Organic Products sector in New Zealand as an application of Veracity in Social and Environmental Audits (SEA) revealed that `Veracity' could have multiple dimensions. These include \emph{regulatory veracity, data veracity, financial data veracity, cultural veracity, and process veracity}. They can be interpreted similarly to the sustainability dimensions in Ojameruaye et al.'s work \cite{ojameruaye2016sustainability}. There can be functional and non-functional software requirements pertaining to these different dimensions of veracity, for example, in a software system that is implemented to audit and certify organic products and its supply chain processes such as manufacturing and distribution \cite{jper_usecasebriefs}. Below, we list potential categories of software requirements that could help satisfy the different dimensions of veracity for the end users in the context of our use case. The non-satisfaction (or the inadequate satisfaction) of these veracity requirements, leading to building the wrong product in terms of the desired level of veracity, can be considered Veracity Requirements TD (VRTD). %We can draw parallels with sustainability dimensions Sustainability dimensions e.g., technical, environmental, social, economic, human

\begin{itemize}
    \item Regulatory Veracity: refers to the compliance with regulations. 
    \begin{itemize}
        \item Compliance Requirements, for example, to comply with the Organic Regulations, Standards, and other regional, local, or international regulations pertaining to different export markets. %e.g., compliance to standards, regional or local or international regulations. 
    \end{itemize}    
    \item Data Veracity: refers to the veracity-related properties of the data entered into the software system such as provenance, transparency, sovereignty, integrity, and accuracy or correctness.  
    \begin{itemize}
        \item Data Provenance or Transparency Requirements (i.e., transparency of the origin of data) --- For example, it should be possible to verify that the results of a particular soil test are actually coming from testing the soil of the particular vineyard that uploads the test results to the certification system.  
        \item Data Sovereignty Requirements (i.e., ownership of data) --- For our use case in New Zealand, the data may be owned by Indigenous communities and require that their identification is not lost when the data is entered into the certification system.
        \item Data Integrity Requirements --- For example, the truthfulness of the on-site audit data must be assured by implementing validation mechanisms and triangulation.
        %\item Data Accuracy Requirements, for example, the data entered in the certification system must be accurate.
        \item Financial Data Veracity Requirements (i.e., the correctness of banking, tax, sales, and accounting data) --- For example, records of inventory management must match with the inventory.
    \end{itemize}
    \item Process Veracity: refers to the adherence to internal organizational policies, processes, methods, and practices. 
        \begin{itemize}
        \item Requirements to support the adherence to the processes. An example from our use case is the practice of having prior approval from the organic certification agency to make certain changes to the Organic management plan (OMP). 
    \end{itemize}   
    \item Cultural Veracity: refers to the trust within the society and culture (human interactions), trust for a brand.
        \begin{itemize}
        \item Requirements that support cultural aspects that may pertain to the trust of Indigenous communities in New Zealand, for example, ensuring that information relating to provenance and cultural authorities is maintained in the digital records associated with the software systems.
        \item Requirements to support the trust for a brand, for example, the issuing of the certification logo for organic produce. 
    \end{itemize}    
\end{itemize}
%refreing back to Ernst's initial definition of RTD i.e.,  the distance between the optimal requirements specification and the actual system implementation, under domain assumptions and constraints.

\subsubsection{`Veracity' as Functional Requirements}

For the above use case, there could be veracity requirements that can be treated as functional requirements. For example, a checklist may need to be implemented as a feature for auditors who may perform site visits on farms and manufacturing facilities to audit the integrity of the data entered into the software system by the farmers. Also, compliance with regulations and organic standards may need specific functions to be implemented in the software system, such as the ability to measure the levels of chemicals in the soil, i.e., the functionality to support soil testing or alternatively, the functionality to import data from other soil testing applications. Furthermore, some checks may be implemented to verify that the data that is entered in the certification system actually belongs to the source of the data, e.g., the farm or vineyard where it comes from.

\subsubsection{`Veracity' as Non-functional Requirements}
\label{non-func-example}
% ASRs, SW quality attributes
'Veracity' can also be treated similarly to any other non-functional requirement, such as reliability, availability, and performance, i.e., similar to a software quality attribute. The prioritized set of requirements or design decisions that do not meet the 'desired level of veracity,' for the software system, i.e., the veracity goals of the system, can be considered Veracity RTD in this case. The desired level of veracity will be measured similarly to measuring the desirability of quality attributes, such as performance or reliability, at the system architecture or data infrastructure level (i.e., how the data is stored, accessed, maintained, and shared), for example, for data veracity. Cultural veracity requirements may also be measured as a non-functional requirement, for example, if the system can be trusted by indigenous communities (for the above use case).

\section{Our Perspective Towards Quantifying Requirements TD for Software Requirements concerning Veracity}
\label{perspective}

%\subsection{Veracity Requirements Technical Debt (VRTD) and its Quantification}

As mentioned before, `Veracity requirements' refers to a specific category of software requirements related to the truth, trust, authenticity, and demonstrability in software-intensive systems. 

Our perspective is that veracity requirements could be seen as functional or non-functional requirements depending on the system under development and the context (e.g., domain constraints, system environment, and business goals), and therefore, RTD for veracity requirements could be quantified by quantifying functional RTD and non-functional RTD. 

Perera et al.'s work \cite{perera2023quantifying} models the concepts that are required for quantifying functional RTD. Their work is grounded in the literature found in their SMS. The non-functional RTD quantification was integrated into their conceptual model in a more recent work \cite{rej_perera}. We posit that non-functional requirements can also lead to design and architectural decisions that may be sub-optimal, and this can be interpreted as non-functional RTD (potentially also as Design TD or Architectural TD). However, for this paper, we stay within the boundaries of RTD. 

%The work of Ojameruaye et al. \cite{ojameruaye2016sustainability, ojameruaye2021sustainability} aligns with our thinking although they discuss sustainability requirements. Therefore, we draw parallels from the work of Ojameruaye et al., \cite{ojameruaye2021sustainability} for veracity in the case of non-functional requirements and provide an illustrative example for quantifying Veracity Debt in Section \ref{example_app} below.

% add the RTD model ? adopt it to VRTD ?

\subsection{An Illustrative Example for an Approach to Quantify Veracity RTD to Support Software Architecture Decision-Making (A Proposal based on Ojameruaye et al. \cite{ojameruaye2016sustainability})}
\label{example_app}

The work of Ojameruaye et al. \cite{ojameruaye2016sustainability, ojameruaye2021sustainability} aligns with our thinking, although they discuss sustainability requirements. Therefore, we draw parallels from the work of Ojameruaye et al. \cite{ojameruaye2021sustainability} for veracity in the case of non-functional requirements and provide an illustrative example for quantifying Veracity Debt below.

Ojameruaye et al. \cite{ojameruaye2016sustainability} develop their RTD quantification based on the Cost-benefit Analysis Method (CBAM) \cite{919103} and the Modern Portfolio Theory (MPT) \cite{mangram2013simplified}. They cover most of the RTD quantification concepts illustrated in the RTD Quantification Model (RTDQM) developed by Perera et al. \cite{perera2023quantifying} to model the quantification of RTD conceptually. Hence, we consider the adoption of the steps in Ojameruaye et al.'s  \cite{ojameruaye2016sustainability} work as a quantification approach for decision-making for VRTD in our example use case. The suggested steps are as follows. 

Software practitioners developing software for the organic certification use case can make informed decisions regarding choosing a software architecture (or a data infrastructure) that supports veracity based on the quantification (or estimation) of veracity RTD.

\begin{enumerate}
    \item Choosing Scenarios and Architectural Strategies (AS) applicable for veracity
    \item Assessing benefits of the veracity dimensions, e.g., the extent
to which the AS is likely to contribute and support data veracity
    \item Quantifying the benefits of the ASs
    \item Quantifying the costs and risk implications of ASs, i.e., the cost to implement the AS and the criticality and likelihood of risks associated with ASs
    \item Calculating desirability by analyzing the costs, benefits, and risks associated with alternative decisions to create an architectural design using MPT thinking to determine how well the desired level of veracity is satisfied by the selection of ASs.  
    \item Making decisions by prioritizing the valid candidate architectures that maximize expected net benefits and minimize risks on dimensions related to veracity requirements (e.g., regulatory veracity, data veracity, financial data veracity, cultural veracity, and process veracity). This can be treated as a portfolio optimization problem to minimize risk and maximize return subject to cost constraints.  
    \item VRTD analysis: Considering that the software architecture is a portfolio of ASs, where each AS has its risks and benefits on goals and veracity dimensions of interest, we could rank the architectures by considering the amount of VRTD that each candidate architecture solution may incur. The VRTD estimation can become the deciding factor in selecting the architecture that will be implemented. 
\end{enumerate}

\section{Research Roadmap}
\label{roadmap}

In this Section, we outline the research roadmap for our work exploring the quantification of Veracity RTD (VRTD), in short, ``Veracity Debt". %Our aim is to develop a conceptual model that helps researchers and practitioners understand the quantification of Veracity Debt and thereby develop quantification approaches that support decision-making for managing Veracity Debt.

\subsection{Veracity Requirements Technical Debt Quantification Model (VRTDQM)}

%In our initial work [1], we developed a conceptual model based on existing literature to understand the concepts and relationships necessary for quantifying both functional and non-functional RTD. We intend to adapt this model for quantifying VRTD by including dimensions and properties related to veracity.

Our aim is to develop a conceptual model that helps researchers and practitioners understand the quantification of Veracity Debt and thereby develop quantification approaches that support decision-making for managing Veracity Debt.

Our plan is to tailor the Requirements Technical Debt Quantification Model (RTDQM) developed by Perera et al. \cite{perera2023quantifying} for the quantification of VRTD by integrating the dimensions of veracity (\emph{regulatory, data, financial data, cultural, and process}) and any new findings that we will derive in the literature review and the engagements with software practitioners from the industry, outlined in Sections \ref{lit_review} and \ref{engagements}, respectively.

We envision that the conceptual model will help researchers and practitioners better understand the quantification of Veracity Requirements Technical Debt and thereby develop mechanisms to quantify VRTD to support informed decision-making to effectively manage the debt. We provided one such illustrative example of how practitioners may go about quantifying Veracity RTD in Section \ref{example_app}. The conceptual model will expand the possibilities of the development of such approaches by broadening the understanding of RTD quantification for veracity requirements. %We hope to provide recommendations to both researchers and practitioners to achieve this. 

\subsection{Literature Review on Veracity and Software Quality Models}
\label{lit_review}

A literature review is underway to survey the literature to identify attributes that could be categorized under the umbrella term `Veracity' as a potential quality attribute for software systems. 

Furthermore, we want to understand whether the different dimensions of veracity that were identified in our previous work can be mapped to properties of veracity discussed in current software quality models. 

%We also want to understand whether the different dimensions of veracity may need to be satisfied generally by software requirements concerning veracity in any domain. Furthermore, we want to identify relationships between the different dimensions of veracity, for example, if one may influence another. 

The expected outcome is to develop a list of abstract concepts pertaining to veracity (pertaining to the previously identified dimensions and potentially newly identified dimensions and with reference to the RTDQM model developed by Perera et al. \cite{perera2023quantifying}) in the context of software-intensive systems and potential metrics to quantify them so that we can make recommendations to practitioners how to utilize them to prevent and manage veracity debt. 

\subsection{Engaging with Software Practitioners from the Industry}
\label{engagements}

\paragraph{Study 1 --- Surveying practitioners from the industry} We surveyed software practitioners to learn whether they formally or informally quantify RTD  in general and for software requirements pertaining to the different dimensions of veracity and what concepts in the conceptual model developed by Perera et al. \cite{perera2023quantifying} were quantified (either formally or informally) by practitioners in the industry. Furthermore, we wanted to seek practitioners' opinions on what concepts may be helpful to quantify to support making informed decisions for fixing RTD instances for software requirements in general and for software requirements concerning veracity.

The practitioners from the industry who answered our survey performed activities such as Requirements Engineering (RE), software design, and implementation; some of them made architectural design decisions, while the rest made business decisions. 

The survey invitation was sent to the participants on the 9th of February, 2024. We closed the survey on April 23, 2024. There was a total of 82 responses. 52 responses were considered valid. 

Practitioners' perceptions gathered via the survey revealed several interesting findings. We report a few key findings in this paper. The conceptual model developed by Perera et al. \cite{perera2023quantifying} guided the development of the survey instrument. %We limit the reporting of the findings to a minimum here since we will report and discuss the survey findings in a separate publication.

\begin{itemize}
    \item Applications/ Technologies developed by practitioners in industry primarily supported Data Veracity Requirements. %100\% of NZ Respondents vs. 73\% of the Rest of the World.
    \item Cultural Veracity Requirements received more attention in New Zealand compared to the rest of the world. % (80\% of NZ Respondents vs. 37\% of the rest of the World). 
    \item Financial Data Veracity Requirements were the most common type reported in critical incidents with a significant impact on the software project due to problems with veracity requirements. %: 67\% of NZ Respondents, 47\% of the Rest of the World.
    \item Most respondents agreed that it is important to quantify the consequences of not fixing Veracity RTD Items (i.e., Veracity Debt Interest)' for decision-making for managing Veracity Debt.
\end{itemize}

\paragraph{Study 2 --- Interviewing practitioners from the industry} A currently ongoing study is semi-structured interviews with software practitioners. This is a study that follows from the survey. It follows a similar structure to the survey, with the advantage of having deeper conversations with software practitioners. 

This study will allow us to validate model concepts in our conceptual model, VRTDQM, while we develop it so that it fits the needs of the industry --- to support the development of new quantification approaches based on our conceptual model. Conversations with practitioners will also help us identify opportunities where existing tools and techniques could support the quantification and management of RTD, especially for software requirements concerning veracity. 

Through this study, we also want to understand whether the different dimensions of veracity may need to be satisfied generally by software requirements concerning veracity in any domain. Furthermore, we want to identify relationships between the different dimensions (or types) of veracity, for example, if one may influence or cause another. In other words, whether requirements pertaining to one dimension could introduce requirements pertaining to another dimension. This may also indicate whether the consequences of RTD for one type of veracity requirement must be quantified and managed over another.

\subsection{Implications to Researchers and Practitioners}
\label{impl}

Below, we summarize the implications to researchers from our perspective of applying the lens of RTD to software requirements concerning veracity.

\begin{itemize}
    \item Software requirements concerning veracity can be both functional and non-functional requirements depending on the system under development and its context. This was evident from the engagements with stakeholders in the organics domain.
    \item There could be different attributes or specific concepts that need to be considered for veracity debt and metrics that help quantify (or measure or estimate) those concepts in software-intensive systems. We further investigate this in our work outlined in Section \ref{lit_review}.
    \item There could be relationships between the different types of veracity requirements, for example, one type introducing requirements pertaining to another type or one type of requirements acting as a constraint to another type of veracity requirements. We further investigate this in our work outlined in Section \ref{engagements}.
    \item There could also be other methods of quantifying and evaluating Veracity RTD in software systems apart from what we proposed as an example in Section \ref{example_app}. For example, the integration of Options Theory \cite{abad2015using}, and the Analytical Hierarchy Process (AHP) could be potential alternatives to portfolio theory \cite{6225999} for decision-making.   
\end{itemize}

\section{Conclusion}
\label{conclusion}

Our preliminary work has led us to see that software requirements concerning veracity, i.e., veracity requirements, can be functional or non-functional, depending on the software system under development and the context. This paper presents our perspective and research roadmap for conceptualizing the quantification of Veracity Requirements Technical Debt (VRTD). Our work proposes a new research stream that researchers could embark on, given that `Veracity' is becoming a primary concern in future and present-day software systems. 

Furthermore, software practitioners could benefit from this thinking to develop software having an awareness of VRTD (i.e., "veracity-aware software systems") to reduce and prevent associated costs and to provide their customers with software products that meet customer expectations in terms of veracity.

\begin{acks}
This research is funded by the New Zealand Ministry of Business,
Innovation and Employment via The Science for Technological Innovation (SfTI) National Science Challenge Veracity Technology Spearhead. We want to especially thank our Theme Leader, Professor Stephen MacDonell, for his kind support throughout.

%Professor Matthias Galster, Co-director of Veracity Lab and Science Lead of Veracity Spearhead, and 

We want to thank Brendan Hoare, Associate Director of Veracity Lab (https://veracity.wgtn.ac.nz/), for engaging us with stakeholders from the Organics Domain in New Zealand and the stakeholders for all the conversations we have had with them. 

We also want to thank Associate Professor Māui Hudson, Co-director of Veracity Lab, and Ernestynne Walsh, Māori data Service Lead of Nicholson Consulting, for the fruitful discussions around cultural veracity.

Our heartfelt gratitude goes to all the advisers and researchers of Veracity Spearhead for the inspiring conversations we have had with them and for their kind support throughout. 

All these conversations between researchers, stakeholders, and partners have shaped our work to date.
\end{acks}

\bibliographystyle{ACM-Reference-Format}
\bibliography{References}

\appendix

%\section{Research Methods}

%\section{Online Resources}

\end{document}